\documentclass{article}
\usepackage{spconf,amsmath,graphicx}
\usepackage{xcolor}
\usepackage{hyperref}

\usepackage{enumitem}
\setlist{nosep, leftmargin=14pt}

\usepackage{mwe} 

\newcommand{\myparagraph}[1]{\noindent\textbf{#1}}

\title{Topological Analysis of Mouse Brain Vasculature via 3D Light-sheet Microscopy Images}
\name{Jiachen Yao$^1$, Nina Hagemann$^2$, Qiaojie Xiong$^1$, Jianxu Chen$^3$, Dirk M. Hermann$^2$, Chao Chen$^1$}
\address{$ ^1$Stony Brook University, Stony Brook, United States\\ 
$ ^2$University Hospital Essen, Essen, Germany \\ 
$ ^3$Leibniz-Institut f{\"u}r Analytische Wissenschaften - ISAS- e.V., Dortmund, Germany}
\begin{document}
%
\maketitle
\begin{abstract}
Vascular networks play a crucial role in understanding brain functionalities. Brain integrity and function, neuronal activity and plasticity, which are crucial for learning, are actively modulated by their local environments, specifically vascular networks. 
With recent developments in high-resolution 3D light-sheet microscopy imaging together with tissue processing techniques, it becomes feasible to obtain and examine large-scale brain vasculature in mice. To establish a structural foundation for functional study, however, we need advanced image analysis and structural modeling methods. 

Existing works use geometric features such as thickness, tortuosity, etc. However,  geometric features cannot fully capture structural characteristics such as the richness of branches, connectivity, etc. 
In this paper, we study the morphology of brain vasculature through a topological lens. We extract topological features\footnote{Our code is provided at \href{https://github.com/TopoXLab/VesselAnalysis}{https://github.com/TopoXLab/VesselAnalysis}} based on the theory of topological data analysis. Comparing of these robust and multi-scale topological structural features across different brain anatomical structures and between normal and obese populations sheds light on their promising future in studying neurological diseases.
\end{abstract}
\section{Introduction}
\label{sec:intro}
Vascular networks play a crucial role in understanding brain functionalities. Neuronal activity and plasticity, which are crucial for learning, are actively modulated by their local environments, including glial, lymphatic and vascular networks. Despite only comprising 2\% of the body weight, brain consumes 20\% of total body energy, which is supplied through the vascular system~\cite{mergenthaler2013sugar,chen2021energy,dienel2001glucose}. Local hemodynamics (the dynamics of blood flow) in the brain is tightly correlated with the surrounding neuronal activity, continuously providing nutrients and bloodborne factors and eliminates waste metabolites to regulate neuronal activity~\cite{katsimpardi2014vascular,trejo2008effects,iadecola2017neurovascular,attwell2010glial}. The coupling between neuronal activity and blood flow, termed neurovascular coupling, is linked to various physiological and pathological brain conditions. Chronic neurovascular coupling dysfunction has been implicated as a key etiology underlying vascular diseases (e.g., stroke and vascular dementia), neurodegenerative disease (e.g., Alzheimer's disease and Parkinson’s disease), sleep disorders and psychiatric disorders (e.g., autism). Many of these pathological conditions are associated with learning deficits. Thus, it is of great importance to understand the function of brain vasculature in learning.

However, current understanding of hemodynamics in neuroscience is far from sufficient. Existing studies use Functional Magnetic Resonance Imaging (fMRI), but lack the fundamental information on detailed brain vasculature structures and cannot account for the highly variable microvascular structures across individual subjects. We need novel statistical and machine learning structural models of brain vasculature to provide important anatomical substrates for functional and mechanistic studies in both physiological and pathological conditions. With recent developments in high-resolution 3D light-sheet microscopy imaging together with tissue processing techniques, it becomes feasible to obtain and examine large-scale brain vasculature in mice~\cite{bennett2022advances}. To establish a structural foundation for functional study, however, we need advanced image analysis and structural modeling methods. 

Existing works~\cite{spangenberg2023rapid,bumgarner2022open,abdellah2020interactive,todorov2020machine} studied mouse brain vasculature using geometric features such as thickness, tortuosity, etc. However, these geometric features cannot fully capture structural characteristics such as the richness of branches, connectivity, etc. These structural cues can be crucial for understanding the physiology, hemodynamics, and eventually brain functionality. 

In this paper, we propose for the first time a suite of topological features for the analysis of mouse brain vasculatures. Our features are based on the theory of topological data analysis \cite{edelsbrunner2010computational,dey2022computational}. In particular, using the theory of persistent homology~\cite{edelsbrunner2002topological}, we extract topological structural information in a robust and multi-scale manner. A thorough analysis of these features on mouse brain vasculature extracted from light-sheet images shows the promise of these topological features in revealing insights into biological and clinical outcomes.

\section{Dataset Description}
\label{sec:data}

We tested our topological analysis framework on a public dataset\footnote{https://zenodo.org/records/6025935} of 3D lightsheet fluorescence microscopy (LSFM) images of blood vessels from various brains regions from wild-type and obese mice and the corresponding vessel segmentation~\cite{spangenberg2023rapid}. Here, we briefly summarize the image acquisition and segmentation method.  

\begin{figure}
    \centering
    \includegraphics[width=1\linewidth]{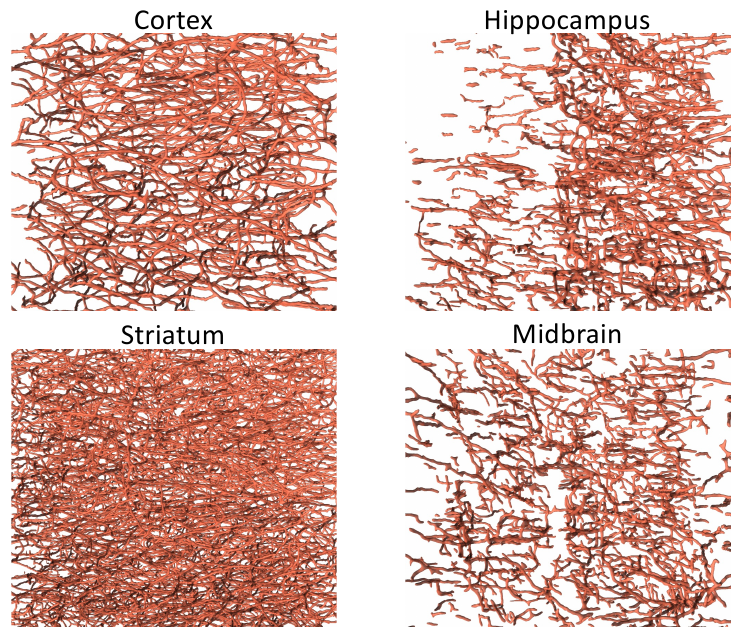}
    \caption{Vessel skeletons at four rostrocaudal levels of the same mouse, observed from a consistent angle, are depicted. As demonstrated, the vessels in the striatum exhibit more complex structures, aligning with the indications from our topological features presented in Fig.\ref{fig:t-test}.}
    \label{fig:vessel}
\end{figure}

First, blood vessels in nine to twelve-week-old male C57BL/6 wild-type mice (group: healthy), or leptin-deficient ob/ob mice (group: obese) of the same age, by perfusing with FITC-albumin-enriched hydrogel. A specific point mutation in the leptin gene of these mice results in the onset of hyperlipidemia, along with related disorders including hyperglycemia, hyperinsulinemia, and infertility. Subsequently, images at various rostrocaudal levels in the brain, such as the striatum, cortex, midbrain, and hippocampus, were acquired with lightsheet fluorescence microscope, UltraMicroscope Blaze™ (LaVision BioTec (Miltenyi), Bielefeld, Germany) with bidirectional light sheet illumination. Finally, the LSFM images were segmented using the image segmentation workflow developed in~\cite{spangenberg2023rapid} and then manually inspected by human experts for quality control. Our experiments in this paper were directly applied on the released segmentation without any changes. In total, the number of images are 12 healthy + 8 obese for cortex, 9 healthy + 5 obese for hippocampus, 6 healthy + 8 obese for midbrain, and 12 healthy + 10 obese for Striatum. An example is shown in Fig.~\ref{fig:vessel}.

\section{Topological Features}
\label{sec:topo-feature}

Our research employs an advanced topological analysis framework that utilizes the Persistent Homology Transform (PHT) to identify and extract distinct topological features from various shapes of vasculature. These extracted features are then rigorously assessed using a permutation test to discern their statistical significance and variability.

\subsection{Persistent Homology}

At the foundation of our framework lies the concept of persistent homology, a pivotal tool in topological data analysis. This method focuses on understanding and characterizing the persistence of specific topological features, such as loops or holes, across different scales of observation. The definition is built on \textit{homology classes} from the classic algebraic topology, which is instrumental in formalizing holes of various dimensions within a topological space. For instance, 0-dimensional holes represent connected components, 1-dimensional holes embody loops, and 2-dimensional holes are akin to voids.

The essence of persistent homology is captured in the process of \textit{filtration}. This involves the sequential construction of a topological shape, where elements are added based on a parameter, such as radius or threshold. As the shape undergoes transformation, we track the persistence of features like holes, enabling a deeper understanding of the topological dynamics of the data.

\begin{figure}
    \centering
    \includegraphics[width=1\linewidth]{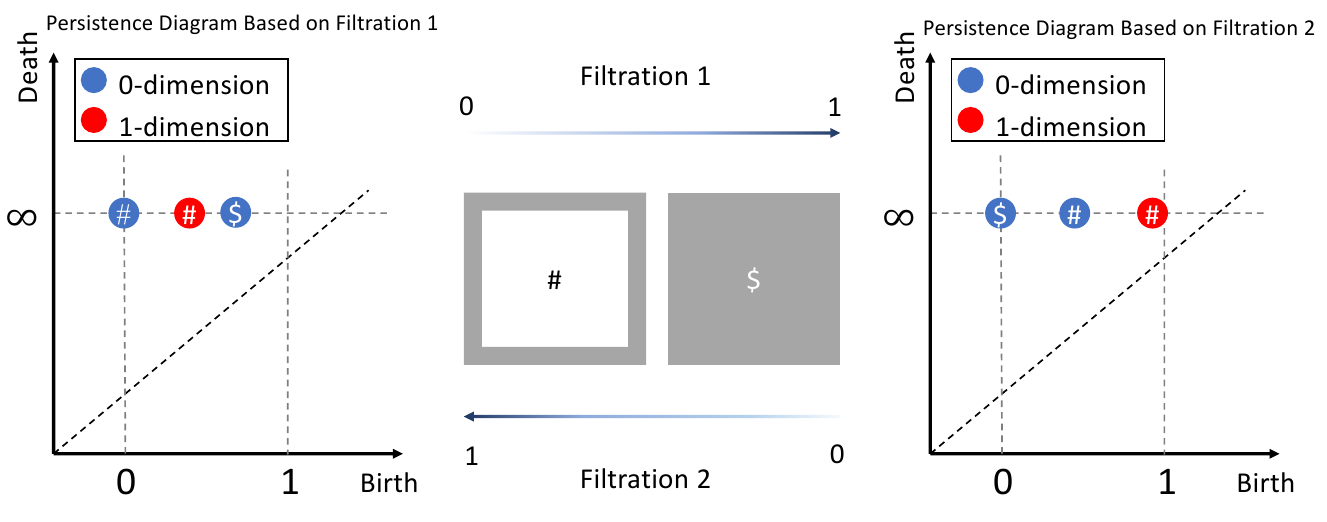}
    \caption{In the case of an object comprising a solid square (marked by \$) and a void square (marked by \#), we apply two distinct filtration functions. These functions are based on the distance from varying directions. Each filtration gives rise to a unique persistence diagram. Collectively, these diagrams from all directions form what is known as a persistence homology transform.}
    \label{fig:PHT}
\end{figure}

For visualization and analysis, we rely on persistence diagrams, as shown in Fig.~\ref{fig:PHT}. These diagrams are pivotal in illustrating the birth and demise times of topological features, thereby shedding light on their persistence throughout the data. In a Cartesian coordinate system, these diagrams plot points above the line y=x, with each point's distance from this line indicating the feature's lifespan—points nearer to the line signify ephemeral features, whereas those further away indicate more enduring characteristics.

\subsection{Persistent Homology Transform}

Our analysis is further enhanced by the implementation of Persistent Homology Transform (PHT), as elucidated in \cite{10.1093/imaiai/iau011}. PHT expands upon the principles of persistent homology, offering a more nuanced analysis of shapes and forms within datasets. This method is particularly advantageous in scenarios where an in-depth understanding of a dataset's geometric structure is essential.

PHT is rooted in persistent homology, focusing on the persistence of data features like holes or loops across diverse scales, ascertained through filtration. Distinctively, PHT employs multiple filtration functions, diverging from traditional approaches that rely on a single filtration perspective. 

This approach in PHT involves applying persistent homology analysis from multiple angles, culminating in a comprehensive depiction of the data. Such a multifaceted representation allows PHT to encapsulate the complete topological nature of a shape or dataset. The integration of varied filtration functions enables PHT to adeptly highlight the complex geometric variations among different shapes and forms, as depicted in Fig.~\ref{fig:PHT}.

Mathematically, PHT is expressed for an object \( M \) as a set of persistence diagrams, formulated as:
\begin{equation}
    PHT(M) = \{PD_{v_i}(M)\}_{i=1}^{k}.
\end{equation}
Here, each persistence diagram \( PD_{v_i} \) is derived using a distinct filtration function based on the distance from different directions \( v_i \), providing a multidimensional perspective on the topological features within the data.

\subsection{Betti Curves and Betti Curve Transform}

To overcome the challenges posed by the multi-point nature of persistence diagrams in machine learning applications, we employ \textit{Betti curves} (BC). These curves effectively convert the complex, multidimensional data from persistence diagrams into a streamlined, one-dimensional vector format. The Betti curve tracks the evolution of topological features (like connected components, holes, and voids) across varying parameters, typically linked to the scale of observation. By observing changes in Betti numbers—which quantify these features—the curve offers a continuous and interpretable topological profile of the data.

Integrating Betti curves with PHT, we compute a Betti curve for each persistence diagram within the PHT framework. This results in a novel feature representation, termed the Betti Curve Transform (BCT):
\begin{equation}
    BCT(M) = \{BC_{v_i}(M)\}_{i=1}^k.
\end{equation}
We utilize the BCT as a key feature in differentiating vasculatures across diverse groups. This approach not only harnesses the topological richness of the data but also adapts it into a format amenable for advanced machine learning analysis, enhancing our ability to discern and categorize the intricate patterns present in vascular structures.

\section{Experiments and Discussion}
\label{sec:exp}

We have applied the proposed methodology to distinguish between obese and healthy mice by analyzing the characteristics of their brain vessels. Our findings demonstrate that the topological features we introduced are effective in detecting differences in the hippocampus and midbrain regions.

\subsection{Hypothesis Test}
\begin{figure}[ht]
    \centering
    \includegraphics[width=0.7\linewidth]{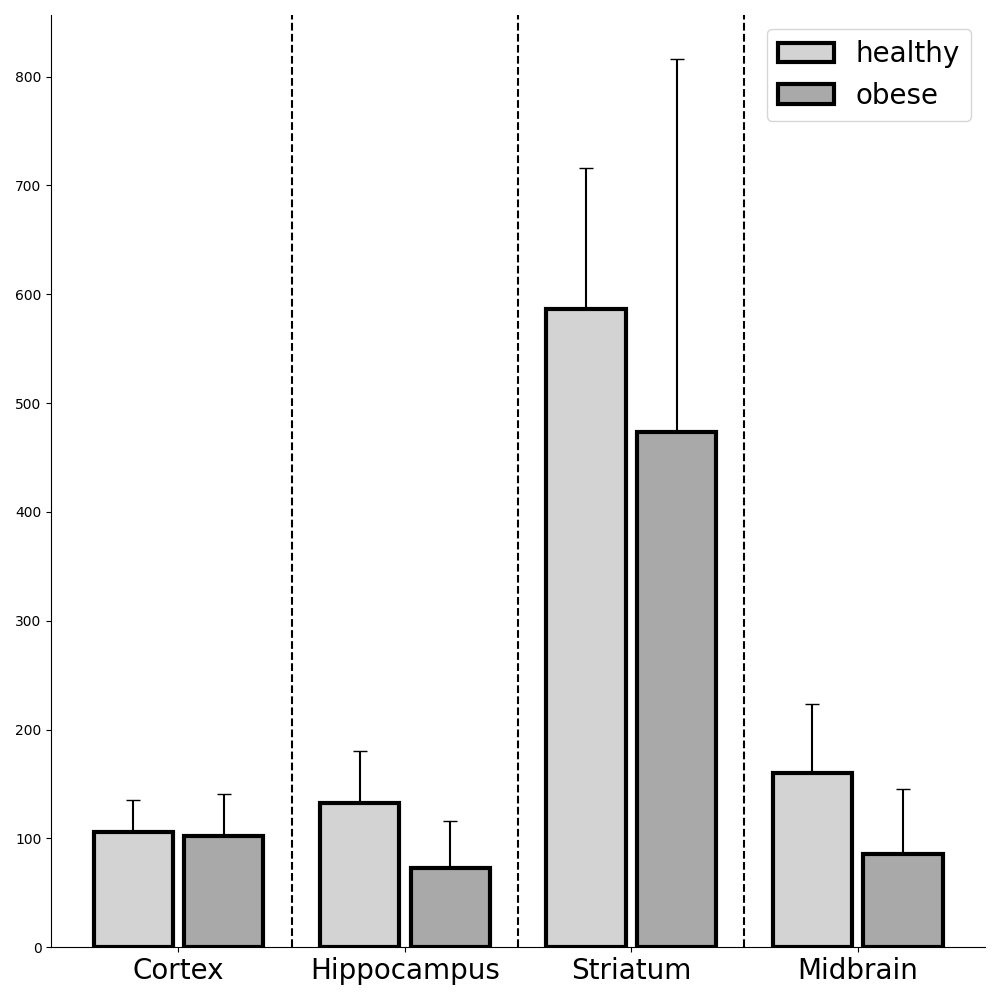}
    \caption{The p-value of a t-test on the mean feature values among each group. For Hippocampus and Midbrain, their p-values are 0.04 and 0.07, respectively. }
    \label{fig:t-test}
\end{figure}

We construct filtration functions from seven different orientations, computing a one-dimensional Betti curve for each respective filtration. These functions are formulated based on the distance from each direction, subsequently normalized to a range of [0, 1]. To discern features between obese and healthy mice, we employ a t-test on the mean feature values , as shown in Fig.~\ref{fig:t-test}. In this figure, a higher value generally signifies an increased number of loops within the vessel, as we are utilizing the 1-dimensional Betti number, which quantifies the count of loops.

\myparagraph{Maximum mean discrepancy.} We also employ a permutation test on the \textit{maximum mean discrepancy} (MMD) statistics. MMD is a statistical measure designed for comparing the distributions of two datasets, proving particularly effective in high-dimensional spaces where conventional statistical tests may be inadequate. MMD functions by assessing the disparity between the mean embeddings of two distributions within a feature space. This space is typically defined through a kernel function, denoted as \( K(\cdot, \cdot) \). Consider two sample sets, \( X \) and \( Y \), comprising \( m \) and \( n \) samples respectively. The MMD formula is expressed as follows:
\begin{align*}
    MMD^2(X,Y) &= \frac{1}{m^2} \sum_{i,j=1}^m K(x_i,x_j) + \frac{1}{n^2} \sum_{i,j=1}^n K(y_i,y_j) \\
    &\quad - \frac{2}{mn} \sum_{i=1}^m\sum_{j=1}^n K(x_i,y_j),
\end{align*}
where \( x_i \) and \( y_j \) represent individual samples from \( X \) and \( Y \) respectively.

To determine the statistical significance of the MMD measure, a permutation test is employed, typically involving \( 3 \times 10^3 \) random permutations. This approach helps in computing the p-value for the MMD statistics, thereby assessing the likelihood of the observed difference occurring by chance. The results, as shown in Table~\ref{tab:MMD}, indicate that the p-value for the midbrain is $0.049$, signifying a statistically significant difference in the vessel structures within this region between obese and healthy mice.

\begin{table}[ht]
    \centering
    \vspace{-.1in}
    \caption{P-value for MMD statistics.}
    \vspace{.1in}
    \begin{tabular}{c|cccc}
    \hline
         & Cortex & Hippocampus & Striatum & Midbrain\\
         \hline
        p-value & 0.258 & 0.411 & 0.506 & 0.049\\
    \hline
    \end{tabular}
    \label{tab:MMD}
\end{table}

\subsection{Compare with Geometric Features}
Our topological features well complements existing geometric features.
In a prior study~\cite{spangenberg2023rapid}, the authors employed statistical analysis on the extracted geometrical features. Within the same experimental conditions, they observed significant differences (p-value $<0.05$) in the vasculature of the striatum and cortex between obese and healthy mice, while no such distinction was noted in the midbrain and hippocampus. In contrast, our analysis reveals that the topological features exhibit significant differences (p-value $< 0.05$) in the vasculature of the hippocampus and midbrain. This comparison suggests a complementary relationship between geometric and topological features.

\section{Conclusion}

Overall, the novel topological analysis using persistent homology presented here provides unprecedented opportunities to gain valuable insights into vascular systems. By labeling of the complete vasculature and comprehensively scanning whole organ samples, we can further investigate the intricate details of recovery processes after events, such as ischemia, in various organs, including the brain and heart. The ability to specifically label certain vascular components such as arteries and veins, in conjunction with the staining of other brain cells such as neurons, astrocytes, pericytes and microglia, provides a solid foundation for understanding network typology. Analyzing complete brain scans also offers a unique opportunity to combine our innovative topological analysis with specific topographical information within distinct brain regions. This approach holds great promise for answering important scientific questions, including elucidating the complexity of brain development in health and disease, evaluating the efficacy of pharmaceutical interventions, and deciphering age-related implications. We expect the developed topological analysis suite paves the road to significantly improve our understanding of vascular dynamics in different physiological and pathological contexts.

\section{Compliance with Ethical Standards}
This research study was conducted retrospectively using microscopy images from animal studies made available in open access in~\cite{spangenberg2023rapid}. Ethical approval was not required as confirmed by the license (Creative Commons Attribution 4.0 International) attached with the original open access data.

\section{Acknowledgement}
The research of J.~Yao and C.~Chen was partially supported by NSF 2144901 and NIH R21CA258493.
The research of J.~Chen was partially supported by the Federal Ministry of Education and Research (Bundesministerium f{\"u}r Bildung und Forschung, BMBF) in Germany under the funding reference 161L0272 and the Ministry of Culture and Science
of the State of North Rhine-Westphalia (Ministerium f{\"u}r Kultur und Wissenschaft des Landes Nordrhein-Westfalen, MKW NRW).

\bibliographystyle{IEEEbib}

\begin{thebibliography}{10}

\bibitem{mergenthaler2013sugar}
Philipp Mergenthaler, Ute Lindauer, Gerald~A Dienel, and Andreas Meisel,
\newblock ``Sugar for the brain: the role of glucose in physiological and pathological brain function,''
\newblock {\em Trends in neurosciences}, vol. 36, no. 10, pp. 587--597, 2013.

\bibitem{chen2021energy}
Yali Chen and Jun Zhang,
\newblock ``How energy supports our brain to yield consciousness: Insights from neuroimaging based on the neuroenergetics hypothesis,''
\newblock {\em Frontiers in Systems Neuroscience}, vol. 15, pp. 648860, 2021.

\bibitem{dienel2001glucose}
Gerald~A Dienel and Leif Hertz,
\newblock ``Glucose and lactate metabolism during brain activation,''
\newblock {\em Journal of neuroscience research}, vol. 66, no. 5, pp. 824--838, 2001.

\bibitem{katsimpardi2014vascular}
Lida Katsimpardi, Nadia~K Litterman, Pamela~A Schein, Christine~M Miller, Francesco~S Loffredo, Gregory~R Wojtkiewicz, John~W Chen, Richard~T Lee, Amy~J Wagers, and Lee~L Rubin,
\newblock ``Vascular and neurogenic rejuvenation of the aging mouse brain by young systemic factors,''
\newblock {\em Science}, vol. 344, no. 6184, pp. 630--634, 2014.

\bibitem{trejo2008effects}
Jos{\'e}~L Trejo, MV~Llorens-Martin, and Ignacio Torres-Alem{\'a}n,
\newblock ``The effects of exercise on spatial learning and anxiety-like behavior are mediated by an igf-i-dependent mechanism related to hippocampal neurogenesis,''
\newblock {\em Molecular and Cellular Neuroscience}, vol. 37, no. 2, pp. 402--411, 2008.

\bibitem{iadecola2017neurovascular}
Costantino Iadecola,
\newblock ``The neurovascular unit coming of age: a journey through neurovascular coupling in health and disease,''
\newblock {\em Neuron}, vol. 96, no. 1, pp. 17--42, 2017.

\bibitem{attwell2010glial}
David Attwell, Alastair~M Buchan, Serge Charpak, Martin Lauritzen, Brian~A MacVicar, and Eric~A Newman,
\newblock ``Glial and neuronal control of brain blood flow,''
\newblock {\em Nature}, vol. 468, no. 7321, pp. 232--243, 2010.

\bibitem{bennett2022advances}
Hannah~C Bennett and Yongsoo Kim,
\newblock ``Advances in studying whole mouse brain vasculature using high-resolution 3d light microscopy imaging,''
\newblock {\em Neurophotonics}, vol. 9, no. 2, pp. 021902--021902, 2022.

\bibitem{spangenberg2023rapid}
Philippa Spangenberg, Nina Hagemann, Anthony Squire, Nils F{\"o}rster, Sascha~D Krau{\ss}, Yachao Qi, Ayan~Mohamud Yusuf, Jing Wang, Anika Gr{\"u}neboom, Lennart Kowitz, et~al.,
\newblock ``Rapid and fully automated blood vasculature analysis in 3d light-sheet image volumes of different organs,''
\newblock {\em Cell Reports Methods}, vol. 3, no. 3, 2023.

\bibitem{bumgarner2022open}
Jacob~R Bumgarner and Randy~J Nelson,
\newblock ``Open-source analysis and visualization of segmented vasculature datasets with vesselvio,''
\newblock {\em Cell reports methods}, vol. 2, no. 4, 2022.

\bibitem{abdellah2020interactive}
Marwan Abdellah, Nadir~Roman Guerrero, Samuel Lapere, Jay~S Coggan, Daniel Keller, Benoit Coste, Snigdha Dagar, Jean-Denis Courcol, Henry Markram, and Felix Sch{\"u}rmann,
\newblock ``Interactive visualization and analysis of morphological skeletons of brain vasculature networks with vessmorphovis,''
\newblock {\em Bioinformatics}, vol. 36, no. Supplement\_1, pp. i534--i541, 2020.

\bibitem{todorov2020machine}
Mihail~Ivilinov Todorov, Johannes~Christian Paetzold, Oliver Schoppe, Giles Tetteh, Suprosanna Shit, Velizar Efremov, Katalin Todorov-V{\"o}lgyi, Marco D{\"u}ring, Martin Dichgans, Marie Piraud, et~al.,
\newblock ``Machine learning analysis of whole mouse brain vasculature,''
\newblock {\em Nature methods}, vol. 17, no. 4, pp. 442--449, 2020.

\bibitem{edelsbrunner2010computational}
Herbert Edelsbrunner and John~L Harer,
\newblock {\em Computational topology: an introduction},
\newblock American Mathematical Society, 2010.

\bibitem{dey2022computational}
Tamal~Krishna Dey and Yusu Wang,
\newblock {\em Computational topology for data analysis},
\newblock Cambridge University Press, 2022.

\bibitem{edelsbrunner2002topological}
Edelsbrunner, Letscher, and Zomorodian,
\newblock ``Topological persistence and simplification,''
\newblock {\em Discrete \& Computational Geometry}, vol. 28, pp. 511--533, 2002.

\bibitem{10.1093/imaiai/iau011}
Katharine Turner, Sayan Mukherjee, and Doug~M. Boyer,
\newblock ``{Persistent homology transform for modeling shapes and surfaces},''
\newblock {\em Information and Inference: A Journal of the IMA}, 2014.

\end{thebibliography}

\end{document}